\documentclass[twocolumn,showpacs,preprintnumbers,amsmath,amssymb]{revtex4}

\usepackage{graphicx}

\begin{document}

\title{Gate tunability of stray-field-induced electron spin precession in a GaAs/In$_x$Ga$_{1-x}$As quantum well below an interdigitated magnetized Fe grating}

\author{L. Meier$^{1,2}$, G. Salis$^1$, C. Ellenberger$^2$, E. Gini$^3$, and K. Ensslin$^2$}
\affiliation{$^1$IBM Research, Zurich Research Laboratory,
S\"aumerstrasse 4, 8803 R\"uschlikon, Switzerland\\
$^2$Solid State Physics Laboratory, ETH Zurich, 8093 Z\"urich,
Switzerland\\
$^3$FIRST Center for Micro- and Nanosciences, ETH Zurich, 8093
Z\"urich, Switzerland}

\date{October 05, 2006}

\begin{abstract}
Time-resolved Faraday rotation is used to measure the coherent
electron spin precession in a GaAs/In$_x$Ga$_(1-x)$As quantum well
below an interdigitated magnetized Fe grating. We show that the
electron spin precession frequency can be modified by applying a
gate voltage of opposite polarity to neighboring bars. A tunability
of the precession frequency of 0.5~GHz/V has been observed.
Modulating the gate potential with a gigahertz frequency allows the
electron spin precession to be controlled on a nanosecond timescale.
\end{abstract}

\maketitle

Considerable effort has been devoted to gaining coherent control
over single electron spins in semiconductors, motivated by the
potentially long coherence times that make such a two-level system
an ideal candidate for a quantum bit~\cite{Loss1998}. Using pulsed
electron spin resonance techniques, Rabi oscillations of single
electron spins have been observed in diamond defect
centers~\cite{Jelezko2004} and in semiconductor quantum
dots~\cite{Koppens2006}. Electrical control of the exchange coupling
between neighboring spins in quantum
dots~\cite{Petta2005,Koppens2005} allows the implementation of
two-qubit gate operations. To address individual spins in an array
of localized spins, either an ac magnetic field has to be applied
locally, or the array has to be exposed to a magnetic-field
gradient, whereby individual spins are addressed by changing the
frequency of a global ac field. The latter approach might be
facilitated by locally tuning the electron g-factor with an electric
field~\cite{Jiang2001,Salis2001}. Also, effective ac magnetic fields
can be provided locally using electric gates, as has been
demonstrated for systems with anisotropic g-factor
tensors~\cite{Kato2003} and for systems with strain-induced
spin-orbit coupling~\cite{KatoNature2004}. Recently, it has been
shown that the magnetic stray field of ferromagnetic structures can
be used to manipulate electron spins~\cite{MeierAPL06}. This idea
has been extended to manipulate single electron spins in a quantum
dot~\cite{Tokura2006} via a spatial displacement in the large and
inhomogeneous magnetic field. Such a spatial displacement can be
induced by applying an electric field to metallic gates, which is
technically easier to achieve than providing an ac magnetic field at
the high frequencies (GHz) involved.

Here, we report on the control of the electron spin precession in a
semiconductor quantum well (QW) using an electric gate voltage and a
magnetic stray field. By employing time-resolved Faraday rotation
(TRFR)~\cite{Crooker1995}, we track the electron spin precession in
a QW below an array of ferromagnetic bars made of Fe. In an external
magnetic field of sufficient strength to magnetize the Fe bars, the
magnetic stray field makes the electron spins precess faster than
below an identical grating made out of non-magnetic
Au~\cite{MeierAPL06}. By applying a gate voltage $V_g$ with opposite
sign to neighboring bars of an interdigitated grating, the electron
distribution in the QW is moved towards the positively charged bar,
and precesses in a higher mean stray field. Application of a voltage
of $V_g = \pm 1$~V to a grating with a period of $1~\mu$m leads to
an increase in the electron spin precession frequency by up to
$0.5$~GHz, corresponding to a magnetic field of ~70~mT. By
modulating the gate voltage with gigahertz frequencies, we achieve
control of the electron precession frequency on the nanosecond
timescale.

Our sample consists of a 40~nm thick InGaAs QW (8.8\% In),
sandwiched between the GaAs substrate and a 20~nm GaAs cap layer.
Both well and cap are $n$-doped with Si to ensure long spin
lifetimes~\cite{Kikkawa1998}, the latter with a $\delta$-doping in
the middle of the layer, the former with a bulk doping aimed at 5
$\times 10^{16} \textrm{cm}^{-3}$. On the surface, arrays of 80~nm
thick Fe (Au) bars have been evaporated using electron-beam
lithography and standard lift-off processes with a 10~nm Ti adhesion
layer between the Fe (Au) and the GaAs surface. The Fe bars were
protected from oxidation by 10~nm Al. Neighboring bars have separate
electrical connections, so that they can be put on different
potentials. We have fabricated interdigitated gratings
100~$\mu$m~$\times$~100~$\mu$m in size having periods $p$ of 1, 2,
and 4~$\mu$m and a bar width of half the period.

We use TRFR to trace the electron spin precession in the QW: a
circularly polarized picosecond pump pulse from a Ti:Sapphire laser
tuned to the absorption edge of the QW ($\lambda$ = 870~nm, P =
500~$\mu$W) and focused ($\approx$20~$\mu$m in diameter) on the
grating creates spin-polarized electrons in the conduction band of
the QW. The polarization axis of the linearly polarized probe pulse
(P = 65~$\mu$W, focused on the same spot), which arrives with a time
delay $\Delta t$ on the sample, is rotated by an angle $\theta$
proportional to the projection of the spin polarization on the laser
beam axis (perpendicularly to the QW). When an external magnetic
field is applied in-plane with the QW and perpendicular to the bars,
the electron spins precess about the magnetic field axis, resulting
in a signal of the form $\theta = \theta_0 \exp{(-\Delta t/T_2^*)}
\cos{(2\pi\nu \Delta t)}$ with the electron spin precession
frequency $\nu = g \mu_B B_\textrm{tot}/h$. Here, $T_2^*$ denotes
the spin lifetime, $g$ the Land\'e g-factor, $\mu_B$ the Bohr
magneton and $B_\textrm{tot} = B_\textrm{ext} + \langle
B_\textrm{stray}\rangle$ the total magnetic field. Experiments were
performed at a temperature of $T=40$~K where effects of nuclear
polarization are below 0.01~GHz~\cite{MeierAPL06}. As our laser
focus is much larger than the grating period, we measure the
electron spin precession averaged over an ensemble of spins that
precess in an inhomogeneous magnetic field. This spatial average is
determined by the laser field distribution below the grating that
acts as an optical mask as well as by the electron distribution
within the illuminated regions of the QW. $\langle
B_\textrm{stray}\rangle$ is the spatially averaged
$B_\textrm{stray}$ that results from these optical and electronic
effects \cite{MeierAPL06}.

\begin{figure}
\includegraphics[width=80mm]{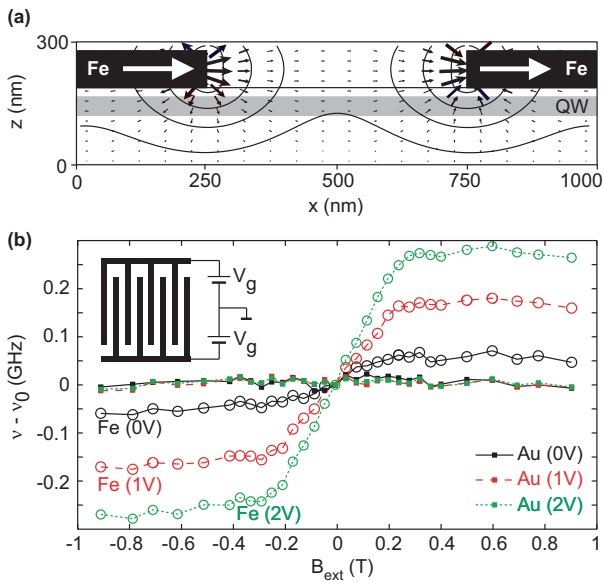}
\caption{\label{fig:fig1}(Color online) (a) Micro-magnetic
simulation of the magnetic stray field of a grating with period
1~$\mu$m. The solid lines of constant field indicate magnetic fields
of 500, 200, 100, and 50~mT. The QW is shaded gray. (b) Electron
precession frequency $\nu$ in the QW below the 1~$\mu$m grating for
different gate voltages. A linear fit $\nu_0 = g \mu_B
B_\textrm{ext}$ to the data has been subtracted (for Fe the fit only
included the saturated region $|B_\textrm{ext}| >$ 0.3~T). Inset:
sketch of a gated interdigitated grating.}
\end{figure}

A numerical simulation of $B_\textrm{stray}$ obtained using the
micro-magnetic simulation tool OOMMF ~\cite{OOMMF} is shown in
Fig.~\ref{fig:fig1}(a). The stray field in the center of the QW is
expected to be $\approx$~200~mT close to a Fe bar and
$\approx$~50~mT in the middle of the gap. The $x$-component, which
our measurement geometry is most sensitive to, changes sign at the
edge of a bar. It is parallel to $B_\textrm{ext}$ between the bars
and antiparallel below a bar.

Figure~\ref{fig:fig1}(b) shows for different $V_g$ the dependence of
$\nu$ on $B_\textrm{ext}$ with a linear background $\nu_0 =
g\mu_BB_\textrm{ext}$ subtracted ($g_\textrm{Au} = 0.5179$,
$g_\textrm{Fe} = 0.5163$). We first focus on the data for $V_g =
0$~V. While on the sample with the Au grating $\nu - \nu_0$ is
constant in $B_\textrm{ext}$, $\nu - \nu_0$ on the Fe sample
increases linearly up to $|B_\textrm{ext}| \approx 0.3$~T. For
$|B_\textrm{ext}| > 0.3$~T, $\nu - \nu_0$ remains constant at about
0.05~GHz, corresponding to an average stray field of $\approx$~7~mT.
Magneto-optical Kerr measurements confirm that at this external
field, the magnetization of the Fe grating (and with it the stray
field) saturates. Simulations assuming a homogeneous electron
distribution and illumination between the bars and no illumination
below the bars predict $\langle B_\textrm{stray}\rangle$ on the
order of $100$~mT. We ascribe the difference to the probing of
negative stray fields antiparallel to $B_\textrm{ext}$ below the Fe
bars owing to optical diffraction at the grating (as $p \sim
\lambda$) and to nonperfect magnetization of the Fe bars due to edge
roughness~\cite{MeierAPL06}.

When applying a voltage of $\pm$~1~V ($\pm$~2~V) to neighboring
bars, the electrons precess 0.15~GHz (0.25~GHz) faster on the Fe
sample than on the Au sample in the saturated region, corresponding
to $\langle B_\textrm{stray} \rangle = $ 20~mT (34~mT). Note that
$\nu - \nu_0$ builds up similarly for all voltage traces. In
particular it saturates at the same value of $B_\textrm{ext}$. This
supports that the same stray field is probed with a different
spatial averaging for different gate voltages. The spin lifetime
$T_2^*$ decreases with an applied gate voltage, also indicating a
change in spatial averaging. However, the situation is more
complicated, since the gate voltage increases the non-radiative
recombination rate as seen in a time-resolved photoluminescence
experiment (data not shown).

\begin{figure}
\includegraphics[width=80mm]{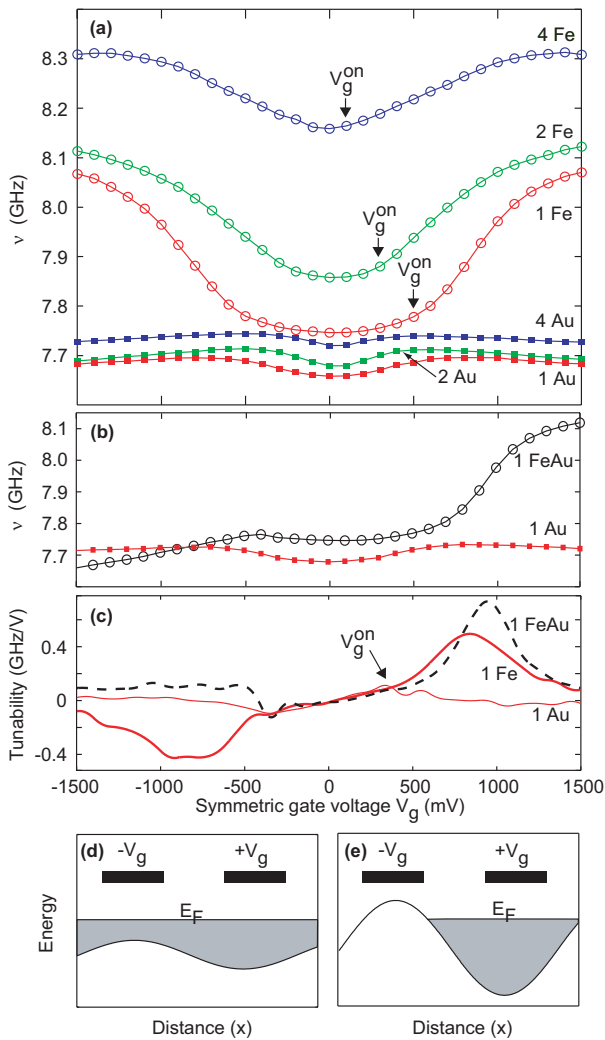}
\caption{\label{fig:fig2}(Color online) (a) Electron precession
frequency $\nu$ at $B_\textrm{ext} = 1.05$~T as a function of $V_g$.
(b) $\nu(V_g)$ for a Au and a mixed Fe/Au grating ($V_g$ is applied
to the Au bars, $-V_g$ to the Fe bars). (c) Gate tunability of the
electron precession frequency $\partial \nu/\partial V_g$ as a
function of $V_g$. (d) and (e) Schematic conduction band modulation
and Fermi energy $E_F$ in the QW for (d) a small and (e) a large
gate voltage $V_g$.}
\end{figure}

In Fig.~\ref{fig:fig2}, $B_\textrm{ext}$ has been set to 1.05~T,
where the magnetization of the Fe bars is saturated.
Figure~\ref{fig:fig2}(a) shows $\nu(V_g)$ for Fe and Au gratings and
three different periods. In samples with Au gratings, $\nu$ changes
little with $V_g$ in contrast to samples with Fe gratings, where
$\nu$ increases by a few tenths of a gigahertz as $|V_g|$ is
increased. We ascribe the small differences in $\nu_\textrm{Au}$ at
$V_g=0$~V for $p=1, 2,$ and $4~\mu$m to small variations in $g$ due
to strain from the grating. Fits yield $g_{Au}^{1,2,4\mu\textrm{m}}
= 0.5211, 0.5225,$ and $0.5253$.

In Fe samples, the increase in $\nu$ with $V_g$ is more pronounced
for gratings with smaller $p$, whereas the stray-field effect on
$\nu$ at $V_g = 0$ is larger for gratings with large $p$. The
understanding of this observation is facilitated by investigating a
mixed Fe/Au grating, in which every other Fe bar has been replaced
by a Au bar. As long as a positive $V_g$ is applied to the Au bars,
$\nu(V_g)$ obtained is similar to $\nu(V_g)$ of the pure Fe sample,
see Fig.~\ref{fig:fig2}(b). However, a negative voltage applied to
the Au bars leads to a \emph{decrease} of $\nu$ on the mixed Fe/Au
sample below the value on the reference Au sample, indicating that
the stray field effectively reduces $B_\textrm{tot}$.

TRFR relies on the circular birefringence at the absorption edge of
the QW. Spin polarization leads to different Fermi energies for
spin-up and spin-down electrons $E_F^{\uparrow, \downarrow}$. The
electron density in our QW is on the order of
$10^{16}~\textrm{m}^{-2}$. With an estimated absorption of 1\% in
the QW, the spin polarization in the conduction band is around 5\%
and $|E_F^\uparrow - E_F^\downarrow| \approx 0.05E_F$ ($\approx
2$~meV). TRFR therefore only probes electrons close to the (mean)
Fermi energy $E_F$.

A positively charged bar leads to an accumulation of electrons below
the bar, but since TRFR only measures electrons close to $E_F$, the
resulting higher electron density does not enhance the TRFR signal.
However, when a bar is negatively charged and all electrons are
depleted from below the bar, then no electrons from this region will
contribute anymore to the TRFR signal and to the averaged $\nu$ [see
Fig.~\ref{fig:fig2}(d) and (e)].

The smaller the grating period $p$, the larger the diffraction that
leads to a probing of negative stray fields below a Fe bar,
resulting in a smaller $\nu$. Besides optical diffraction, effects
of surface-plasmon-enhanced transmission \cite{Ebbesen1998,
Schroter} through the metallic gratings might also possibly play a
role in the illumination of the QW below a bar. Applying $\pm V_g$
to a pair of neighboring bars has no effect on the bar at $+V_g$,
but removes contributions from electron spins in the negative stray
field below the bar at $-V_g$, which results in an overall increase
of $\nu$. The larger the illuminated region below a bar (i.e., for
small $p$), the larger the increase in $\nu$ when this region is
depleted below the negatively charged bar.

Let us now return to the pure Fe grating. The increase in $\nu$ is
not linear in $V_g$. The sensitivity $\partial \nu/\partial V_g$ of
$\nu$ to changes in $V_g$ for the $p=1~\mu$m grating is plotted in
Fig.~\ref{fig:fig2}(c). For $V_g < V_g^{\textrm{on}} \approx 500$~mV
the increase is small and very similar for both the Fe and the Au
(as well as the mixed Fe/Au) grating. We suspect that in this
regime, $\nu$ changes because of a variation of the electron
$g$-factor by about 0.001. Possible explanations for such an
electric-field-induced modification of spin dynamics include changes
in the overlap between electron and hole wavefunctions
\cite{Gerlovin2004}, band-structure effects \cite{Baron2003,
Ivchenko1997}, and strain-induced spin-orbit effects
\cite{Malinowski2000}. When probing a single 2~$\mu$m wide gap
between two large, electrically contacted Fe (Au) gates (data not
shown), we find a very similar behavior in $\nu(V_g)$ as in the case
of the Fe (Au) grating with $p=4~\mu$m (i.e., 2~$\mu$m bar and
2~$\mu$m gap). Hence, measuring a grating is equivalent to measuring
a single gap, but the grating enhances the signal-to-noise ratio
considerably. Specifically, in both cases, the sign of the
$x$-component of the electric field is not of importance: in a
single gap the electric field always points in one direction,
whereas on a grating, we average over fields pointing in the $x$-
and in the $-x$-direction.

For $|V_g| > V_g^{\textrm{on}}$, $\nu$ increases strongly on the
1~$\mu$m Fe sample, whereas it remains constant on the 1~$\mu$m Au
sample. The tunability is highest around $V_g \approx 800$~mV, where
a change of 1~V in $V_g$ leads to a variation of about 0.5~GHz in
$\nu$, corresponding to an effective stray field of $\approx~70$~mT.

The on-set voltage $V_g^{\textrm{on}}$ is smaller for larger $p$. A
straightforward analysis of the electric field between and below the
bars can explain this dependence on $V_g$. As mentioned above, the
relevant mechanism that increases $\nu$ is the depletion of the QW
below the negatively charged gate. In the center below a bar, far
away from an edge, vertical electric fields dominate. With a
capacitor model, we estimate the potential drop between gate and QW
needed to deplete the QW to be roughly 100~mV, which is lower than
$V_g^{\textrm{on}}$. For voltages which are more negative than
required for the depletion of the electron gas, lateral electric
fields between the bars become important, similar to the situation
in quantum point contacts~\cite{vanWeesQPC}.

Close to the edge of a bar, where the magnetic stray fields are
highest, however, lateral electric fields dominate. Under
illumination at $V_g = 1$~V, a considerable current of approximately
$1~\mu$A is measured through a gated grating, yielding an estimated
resistance of 2~M$\Omega$. This resistance can be seen as a series
of three resistances: a forward-biased Schottky barrier $R_s^f$ from
one bar to the QW, the resistance $R_\textrm{QW}$ of the QW itself,
and a reverse-biased Schottky barrier $R_s^r$ from the QW to the
other bar. The period $p = d_{gate} + d_{gap}$ is the sum of the
gate and the gap width. The (lateral) electric field in the QW is
\begin{eqnarray*}
    E_\textrm{QW} = \frac{V_\textrm{QW}}{d_{\textrm{gap}}} & = & \frac{R_\textrm{QW}}{d_{\textrm{gap}}} \frac{2V_g}{R_s^f + R_s^b +
    R_\textrm{QW}} \\ & \approx & \frac{V_g}{d_{\textrm{gap}}} \frac{R_\textrm{QW}}{R_{s}^f +
R_s^b},
\end{eqnarray*}
where we have assumed $R_{s}^b \gg R_\textrm{QW}$. Enlarging the
channel length $d_{\textrm{gap}}$ increases its resistance, i.e.
$R_\textrm{QW} \propto d_{\textrm{gap}}$, whereas enlarging
$d_{\textrm{gate}}$ reduces the resistance of the Schottky contact,
as the area between gate and the sample surface is increased, thus
$R_s^{f,b} \propto 1/d_{\textrm{gate}}$. As a consequence,
$E_\textrm{QW} \propto d_{\textrm{gate}} V_g$.

We assume that a critical (lateral) field
$E_\textrm{QW}^\textrm{on}$ is needed to significantly shift the
depletion edge of the electron gas close to the bar edge and to move
the electrons toward the positively charged bar. Then, for larger
$d_{\textrm{gate}}$ the onset voltage decreases:
\begin{equation}\label{Vgon}
    V_g^{\textrm{on}} \propto
    E_\textrm{QW}^\textrm{on}/d_{\textrm{gate}}.
\end{equation}
This qualitatively explains the dependence of $V_g^\textrm{on}$ on
the grating geometry. In addition, we experimentally tested relation
(\ref{Vgon}) by fabricating gratings with $d_\textrm{gap} = a
d_\textrm{gate}$, a = 1, 2, and 3. Changing $d_\textrm{gap}$ did not
significantly alter $V_g^{\textrm{on}}$, whereas a larger
$d_\textrm{gate}$ reduced $V_g^{\textrm{on}}$ substantially.

\begin{figure}[hbt]
\includegraphics[width=80mm]{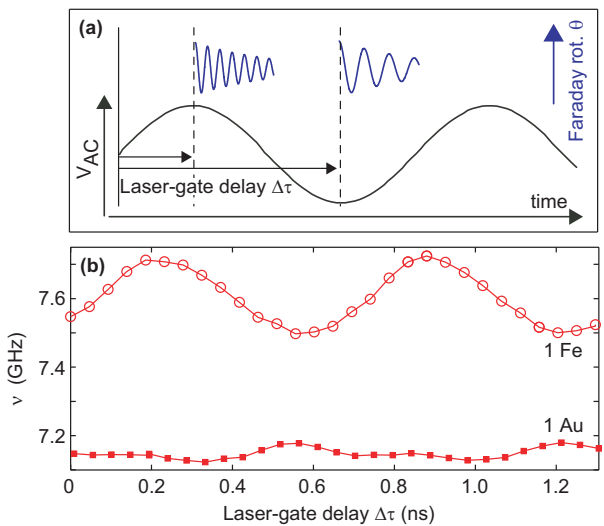}
\caption{\label{fig:fig3}(Color online) (a) Measurement of the
electron spin precession at different phase shifts between laser and
ac voltage (see text for details). (b) Electron precession frequency
$\nu$ as a function of the phase shift between laser pulse and ac
voltage phase.}
\end{figure}

In Fig.~\ref{fig:fig3} we present results with a
high-frequency-modulated gate voltage on the $p=1~\mu$m grating.
While all odd bars are put on ground potential, we apply
$V_\textrm{dc} = -2$~V [corresponding to $V_g = -1$~V in
Fig.~\ref{fig:fig2}(a)] to the even bars and add a modulation
$V_\textrm{ac}(t) = V_\textrm{ac}^0\sin{\left( 2\pi f t \right)}$,
with $f = 1.44$~GHz. An RF power of 10~dBm was used, corresponding
to a $V_\textrm{ac}^0$ of about 0.7~V. Figure~\ref{fig:fig3}(a)
explains the measurement principle: by scanning the phase difference
between the laser pulse and the ac modulation $\Delta\tau$, we are
able to track the electron spin precession at different phases of
the ac modulation.

Figure~\ref{fig:fig3}(b) shows $\nu$ as a function of $\Delta\tau$.
Whereas on the Au sample $\nu$ oscillates only weakly with
$\Delta\tau$, $\nu$ exhibits a strong periodic behavior with a
period of $1/f$ and an amplitude $\nu_1$ of about 0.1~GHz on the Fe
sample. This is explained by assuming that the electrons in the QW
follow the ac modulation. When $\Delta\tau$ is such that the laser
pulse coincides with a minimum in $V_\textrm{ac}$, then the voltage
below the even bars is negative most of the time we measure the
electron spin precession, depleting the electrons in the QW below
and, as explained for the dc case above, $\nu$ is maximal. A similar
argument explains the minima in $\nu$. For experimental reasons, we
cannot quote $\Delta\tau$ in absolute numbers, we measure only the
relative change.

A quantitative estimate for the oscillation amplitude $\nu_1$ can be
obtained from the derivative of $\nu(V_g)$, given in
Fig.~\ref{fig:fig2}(c). Taking into account that only one gate is
modulated, we estimate
\begin{equation*}
  \nu_1 = \frac{1}{2} \left.\frac{\partial \nu}{\partial V_g}\right|_{V_g = -1~\textrm{V}} V_\textrm{ac}^0
  \approx
  0.18~\textrm{GHz,}
\end{equation*}
which is in reasonable agreement with our measurement. To prevent
the effects of $V_\textrm{ac}(t)$ on $\nu$ from being averaged out,
we extract $\nu$ only from a few oscillations during the first
$T_\textrm{fit} = 400$~ps of the spin precession (shorter than the
ac period $T_\textrm{ac} = 1/f \approx 700$~ps). Since a few
($\approx$ 3) electron spin oscillations are needed to determine
$\nu$ with ample precision, we cannot fulfill the ideal case of
$T_\textrm{fit} \ll T_\textrm{ac}$, leading to some reduction of the
observed oscillation amplitude of $\nu$. At lower frequencies on the
order of several 100~MHz, we have measured an oscillating
$\nu(\Delta\tau)$ also on the Au sample that cannot be explained by
a magnetic stray field and will be subject of further
investigations.

In conclusion, we have observed a tunability of the electron spin
precession below a magnetized Fe grating. By applying a gate voltage
to an interdigitated grating, we are able to tune the electron spin
precession frequency by 0.5~GHz/V, corresponding to an effective
magnetic stray field of about 70~mT/V. Modulating the gate potential
with a frequency of 1.44~GHz enables the electron spin precession to
be controlled on a nanosecond timescale. This could be useful for
further experiments that address electron spin resonance using
magnetic stray fields.

We thank O. Homan for help with sample preparation, M. Tschudy for
the evaporation of Fe, and R. Allenspach and T. Ihn for fruitful
discussions.

\end{document}